\documentclass{IEEEcsmag}

\usepackage[colorlinks,urlcolor=blue,linkcolor=blue,citecolor=blue]{hyperref}
\usepackage{tikz}

\newcommand*\circled[1]{\tikz[baseline=(char.base)]{
            \node[shape=circle,draw,inner sep=0pt,fill=black, text=white] (char) {#1};}}
\newcommand{\myGlobalTransformation}[2]
{
    \pgftransformcm{1}{0}{0.4}{0.5}{\pgfpoint{#1cm}{#2cm}}
}

\newcommand{\gridThreeD}[3]
{
    \begin{scope}
        \myGlobalTransformation{#1}{#2};
        \draw [#3] grid (5,5);
    \end{scope}
}

\newcommand{\gridThreeDSecond}[3]
{
    \begin{scope}
        \myGlobalTransformation{#1}{#2};
        \draw [#3] grid (3,3);
    \end{scope}
}

\newcommand{\gridThreeDThird}[3]
{
    \begin{scope}
        \myGlobalTransformation{#1}{#2};
        \draw [#3] grid (1,1);
    \end{scope}
}

\tikzstyle myBG=[line width=3pt,opacity=1.0]

\newcommand{\dotesUpper}[2]
{
    \begin{scope}
        \myGlobalTransformation{#1}{#2};
   
     \node at (2.5,0.5) [circle,fill=gray] {};
     \node at (1.5,1.5) [circle,fill=gray] {};
    \node at (3.5,1.5) [circle,fill=gray] {};

     \node at (0.5,2.5) [circle,fill=gray] {};
      \node at (2.5,2.5) [circle,fill=gray] {};
      \node at (4.5,2.5) [circle,fill=gray] {};
      \node at (1.5,3.5) [circle,fill=gray] {};
         \node at (3.5,3.5) [circle,fill=gray] {};
           \node at (2.5,4.5) [circle,fill=gray] {};

    \end{scope}
}

\newcommand{\lowerDots}[2]
{
    \begin{scope}
        \myGlobalTransformation{#1}{#2};
 
     \node at (0.5,1.5) [circle,fill=gray] {};
       \node at (-1.5,1.5) [circle,fill=gray] {};
          \node at (-0.5,0.5) [circle,fill=gray] {};
           \node at (-0.5,2.5) [circle,fill=gray] {};

    \end{scope}
}

\newcommand{\lowestDot}[2]
{
    \begin{scope}
        \myGlobalTransformation{#1}{#2};
 
    \node at (0.5,1.5) [circle,fill=black] {};

    \end{scope}
}
\usepackage{amsmath}
\usepackage{upmath}
\usepackage{mathtools}
\usepackage{adjustbox}
\usepackage{cite}
\usepackage[font=small,labelfont=bf,textfont={bf},tableposition=top]{caption}
\usepackage{xspace}
\usepackage{floatrow}
\usepackage[labelfont=bf,textfont=normalfont,singlelinecheck=off,justification=raggedright]{subcaption}
\usepackage{xurl}
\usepackage{amsmath,amssymb,amsthm}

\usepackage{xparse}
\usepackage{upmath}
\usepackage[italic]{mathastext}
\usepackage{stfloats}
\usepackage{comment}

\usepackage[resetlabels,labeled]{multibib}
\newcites{T}{REFERENCES}
\usepackage{setspace}
\usepackage{ragged2e}

\usepackage[colorinlistoftodos, textwidth=\marginparwidth]{todonotes}

\let\labelindent\relax
\usepackage{enumitem}
\usepackage{paracol}
\usepackage[many]{tcolorbox}
\usepackage{lipsum}

\newtcolorbox{NewBox}[1]{%
  floatplacement={#1}, width=0.625\textwidth,
  colframe=gray!10!black,colback=orange!10!white,boxrule=1pt,arc=.2em,boxsep=-1.6mm,
  }

\usepackage[pscoord]{eso-pic}

\newcommand{\placetextbox}[3]{
  \setbox0=\hbox{#3}
  \AddToShipoutPictureFG*{
    \put(\LenToUnit{#1\paperwidth},\LenToUnit{#2\paperheight}){\vtop{{\null}\makebox[0pt][c]{#3}}}%
  }%
}%

\newcommand{\juan}[1]{{\color{black}#1}}

\newcommand{\gagan}[1]{{\color{black}#1}}

\newcommand{\gmicro}[1]{{\color{black}#1}}
\newcommand{\gagann}[1]{{\color{black}#1}}
\newcommand{\damla}[1]{{\color{black}#1}}

\newcommand{\damlaa}[1]{{\color{black}#1}}
\newcommand{\sr}[1]{{\color{black}#1}}
\newcommand{\src}[1]{{\color{black}#1}}
\newcommand{\did}[1]{{\color{black}#1}}
\newcommand{\juang}[1]{{\color{black}#1}}
\newcommand{\gmicroF}[1]{{\color{black}#1}}
\newcommand{\damlaaa}[1]{{\color{black}#1}}
\newcommand{\alser}[1]{{\color{black}#1}}
\newcommand{\aov}[1]{{\color{black}#1}}
\newcommand{\sneaky}{\texttt{SneakySnake}\xspace}
\newcommand{\vadvc}{\texttt{vadvc}\xspace} 
\newcommand{\hdiff}{\texttt{hdiff}\xspace} 

\newcommand{\gcamera}[1]{{\color{black}#1}}
\newcommand{\gonur}[1]{{\color{black}#1}}
\newcommand{\gonurnew}[1]{{\color{black}#1}}
\newcommand{\Hl}[2][\empty]{%
\ifx#1\empty
\else
\sethlcolor{#1}%
\fi
\hl{#2}}
\usepackage{soul,color}
\soulregister\Hl{7}
\soulregister\ref7
\soulregister\cite7
\soulregister\pageref7
\newcommand{\garxiv}[1]{{{#1}}}

\newcommand{\garxivCite}[1]{{#1}}

  \newcommand{\MYhref}[3][blue]{\href{#2}{\color{#1}{#3}}}%

\newcommand{\affilCMU}{$^{\Join}$}

\newcommand{\affilIBM}{$^{\triangledown}$}

\newcommand{\affilETH}{$^{\diamond}$}

\newcommand{\affilTUE}{$^{\star}$} 

\jvol{40}
\jnum{5}
\paper{8}
\jname{IEEE MICRO}

\usepackage{floatrow}
\begin{document}



\placetextbox{0.55}{0.87}{\textsf{\emph{This is an extended and updated version of a paper published in}}}%
\placetextbox{0.55}{0.85}{\textsf{\emph{IEEE Micro, vol. 41, no. 4, pp. 39-48, 1 July-Aug. 2021}}}%
\placetextbox{0.55}{0.83}{\textsf{\emph{\url{https://doi.org/10.1109/MM.2021.3088396}}}}%

\title{
\vspace{27pt}
\begin{center}\textbf{
\scalebox{2}{\fontsize{11.4pt}{0pt}\selectfont \textbf{FPGA-Based Near-\gagann{Memory} 
Acceleration}}
\scalebox{2}{\fontsize{11.4pt}{0pt}\selectfont \textbf{of Modern Data-Intensive Applications}}
} 
\end{center}}

\author{
\vspace{-30pt}
\large
\begin{center}
{Gagandeep Singh\affilETH}\quad%
{Mohammed Alser\affilETH}\quad%
{Damla Senol Cali\affilCMU}\quad\\
\vspace{5pt}
{Dionysios Diamantopoulos\affilIBM}\quad
{Juan G{\'{o}}mez-Luna\affilETH}\quad\\
\vspace{5pt}
{Henk Corporaal\affilTUE}\quad%
{Onur Mutlu\affilETH\affilCMU}\\%
\end{center}
}
\affil{\normalsize
\begin{center}
\it\affilETH ETH Z{\"u}rich \quad 
\affilCMU Carnegie Mellon University\\
\vspace{2pt}
\affilTUE Eindhoven University of Technology
\quad \affilIBM IBM Research Europe   \\
\vspace{-25pt}
\end{center}
}

\begin{abstract}
\justifying
\gagan{Modern data-intensive applications demand high comput\gonur{ation} capabilities with strict power constraints. Unfortunately, such applications suffer from a significant waste of both execution cycles and energy in current computing systems due to the costly data movement \gagan{between the comput\gonur{ation} units and the memory units}. \damla{Genome analysis} and weather prediction are \sr{two examples of such applications}. \sr{Recent FPGAs} couple \gonur{a} reconfigurable fabric with high-bandwidth memory (HBM) \sr{to enable more \gonur{efficient} data movement and improve overall performance and energy efficiency}. This trend \gonur{is an example of} a paradigm shift to \emph{near-memory \gcamera{computing}}. \sr{We leverage such} an FPGA \sr{with \gonur{ high-bandwidth memory (HBM) for improving}} \damla{the pre-alignment filtering step of genome analysis and} representative kernels from a weather prediction model.  Our evaluation \gonur{demonstrates} large speedups and energy savings \gonur{over} a high-end IBM POWER9 system and a conventional FPGA board with DDR4 memory. We conclude that \damlaa{FPGA-based} near-memory \gcamera{computing} has the potential to \gonur{alleviate} the data movement bottleneck for modern data-intensive applications.}
\vspace{-5pt}
\end{abstract}

\maketitle

\chapterinitial{Modern computing systems} \damlaa{suffer} from \gmicro{a} large gap between the performance and energy efficiency of comput\gonur{ation} units and \gonur{memory} units.
\gagan{These systems follow a \emph{processor-centric} approach where data has to move \sr{back and forth} from the memory units using a relatively slow and power-hungry off-chip bus to the comput\gonur{ation} units for processing.}
\gagan{As a result}, data-intensive \damlaa{workloads} (e.g., \gcamera{genome analysis}~\garxiv{\cite{nag2019gencache,firtina2020apollo,li2009fast,ren2018efficient,turakhia2018darwin,li2018minimap2,alser2020accelerating,alser2020technology,kim2018grim,senol2019nanopore,senolcalimicro2020,DRAGEN2020,lapierre2020metalign, kim2021airlift, lapierre2019micop}} and weather \gcamera{modeling}~\cite{doms1999nonhydrostatic,thaler2019porting,singh2020nero,singh2019narmada}) require continuous \sr{memory-CPU-memory} data movement, which imposes \gagan{an} extremely large overhead in terms of execution time and energy efficiency~\cite{ mutlu2019processing}.

\begin{figure}[h]
 \centering
 \includegraphics[width=1.04\linewidth,trim={0.3cm 0.2cm 0cm 0.2cm},clip]{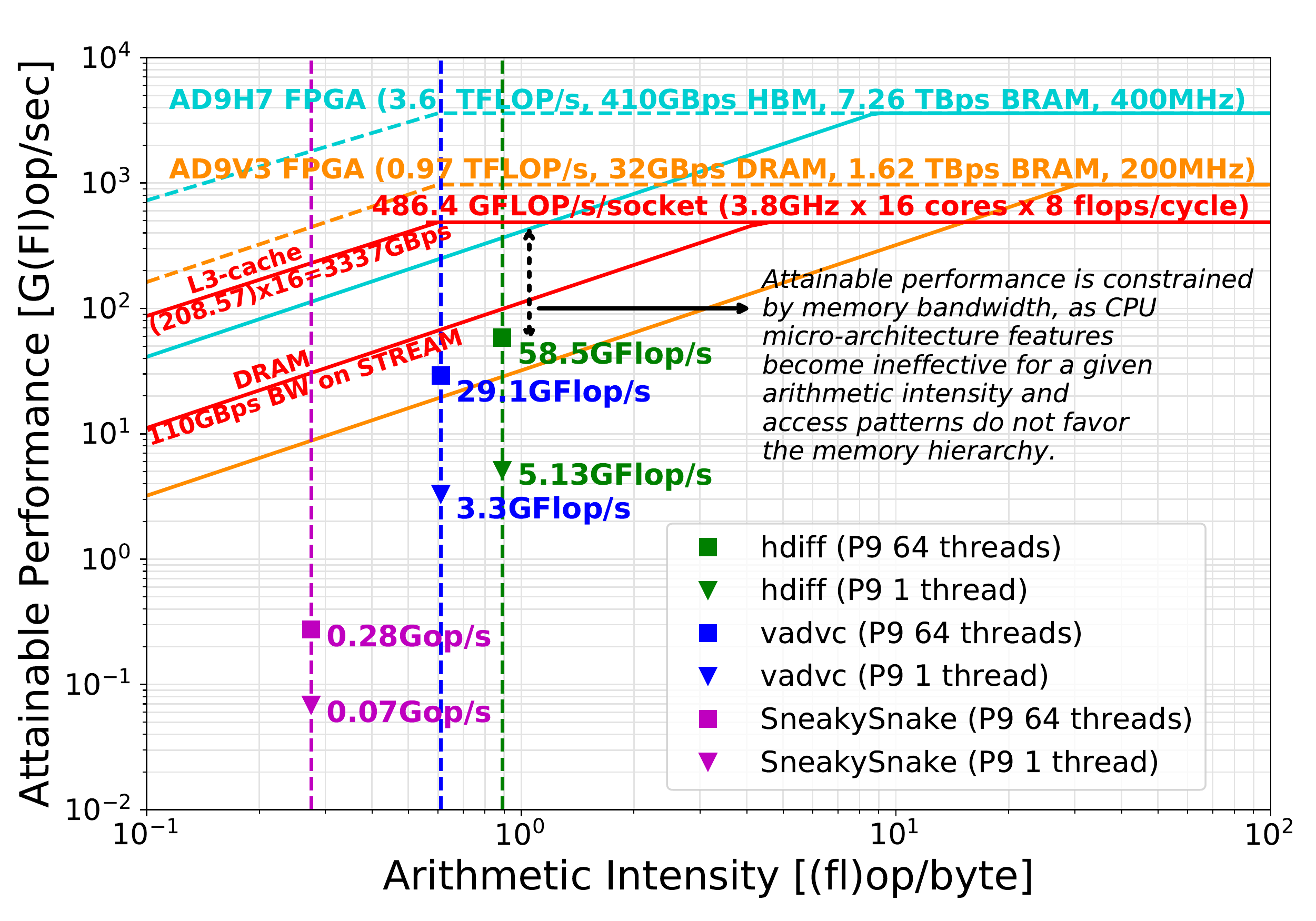}
  \vspace{-15pt}
  \caption{Roofline for POWER9 (1-socket) showing \gagan{pre-aligment filtering algorithm (\sneaky),} \damlaaa{and} vertical advection {(\texttt{vadvc})} and horizontal diffusion {(\texttt{hdiff})} kernels \damlaa{from the COSMO weather prediction model} for single-thread and 64-thread 
  implementations. 
  The plot shows also the rooflines of the FPGAs used in our work \gmicro{with peak DRAM and on-chip BRAM bandwidth.}}
 \label{fig:roofline}
\end{figure}

\sr{We provide in} \juan{Figure~\ref{fig:roofline} the roofline model \garxiv{\cite{williams2009roofline}} 
on an IBM POWER9 CPU \gonurnew{(IC922)} \garxivCite{\cite{sadasivam2017ibm}}  for \damlaa{the state-of-the-art pre-alignment filtering algorithm for} genome analysis~\cite{alser2019sneakysnake} and two compound stencil kernels from the \damlaa{widely-used} COSMO (Consortium for Small-Scale Modeling) weather prediction model~\cite{doms1999nonhydrostatic}. 
\sr{A key observation is that both applications have low arithmetic intensity with complex memory access behavior. The pre-alignment filtering algorithm, \sneaky, builds a special matrix (called a chip maze in Section ``Case Study 1: Pre-Alignment Filtering in Genome Analysi\gcamera{s}'') used to calculate an optimal solution for \gonur{the} pre-alignment filtering problem. \sneaky calculates only portions of this chip maze to maintain \gonur{speed}. This involves irregular visits to different entries of the chip maze, \gonur{leading to} a strong mismatch between the nature of data access patterns and the layout of data in memory for \sneaky. Such a mismatch leads to limited spatial locality \gonur{and cache effectiveness}, causing frequent data movement between the memory subsystem and the processing units.} The weather kernels (\vadvc and \hdiff) consist of compound stencils that perform a series of element-wise computations on a three-dimensional grid~\cite{gysi2015modesto}. Such compound kernels are dominated by DRAM-latency\gonur{-}bound operations due to complex memory access patterns. 
As a result, the performance of these \sr{applications} is significantly lower than the peak CPU performance. \gmicro{This is a common trend in various data-intensive workloads~\garxiv{\cite{oliveira2021pimbench,nair2015active,mutlu2021primer_pim, nai2017graphpim,7056040,lee2018application,kang2013enabling,hashemi2016continuous,akin2015data,babarinsa2015jafar,lee2015bssync,chi2016prime,kim2016neurocube,asghari2016chameleon,boroumand2016lazypim,seshadri2015gather,liu2017concurrent,gao2015practical,morad2015gp,googleWorkloads,teserract,ahn2015pim,hsieh2016accelerating,hashemi2016accelerating,singh2019review,fernandez2020natsa,hajinazar2021simdram,NAPEL,upmem2021}}.}}
  
\damla{In this work, our} \textbf{goal} is to overcome the memory bottleneck of two key real-world \damlaa{data-intensive} applications\damla{, genome analysis and weather \gcamera{modeling},} by exploiting near-memory computation capability on modern FPGA accelerators with high-bandwidth memory (HBM)~\garxivCite{\cite{hbm}} that are attached to \gonur{a} host CPU.
The use of FPGAs can yield significant performance improvements, especially for parallel algorithms. 
\gagan{Modern FPGAs \damlaa{provide} four key trends:
\begin{enumerate} 
\item The integration of high-bandwidth memory (HBM) on the same package \damlaa{with} an FPGA allows us to implement our accelerator logic much closer to the memory with \damlaa{an order} of magnitude more bandwidth than traditional DDR4-based FPGA boards. Thus, these modern FPGAs adopt a \gonur{more} \emph{data-centric} approach \gonur{to computing}. 

\item FPGA manufacturers have introduced UltraRAM (URAM)~\garxiv{\cite{uram}} along with the Block RAM (BRAM) that offers \sr{\damlaa{large} on-chip memory next to the logic}.

\item Recent FPGA boards with new cache-coherent interconnects (e.g., IBM Coherent Accelerator Processor Interface (CAPI)~\garxivCite{\cite{stuecheli2015capi}}, Cache Coherent Interconnect for Accelerators (CCIX)~\garxiv{\cite{benton2017ccix}}, and Compute Express Link (CXL)~\garxiv{\cite{sharma2019compute}}) employ a shared memory space that \src{allows} tight integration of FPGAs with CPUs at high bidirectional bandwidth (on the order of tens of GB/s). This integration allows \gonur{the FPGA} to \did{coherently} access the host system's memory \gonur{using} a pointer, rather than \gonur{requiring} multiple copies of the data. 

\item \damlaa{Newer FPGAs} are manufactured with an advanced technology node of 7-14nm FinFET~\garxivCite{\cite{gaide2019xilinx,vu37p}} \sr{that} offers higher performance.
\end{enumerate}
\sr{These four} trends suggest that \sr{modern FPGA architectures deliver unprecedented levels of integration and compute capability due to new \damlaa{advances and features}, which} provides an opportunity to \gonur{largely alleviate} \gagan{the} \emph{memory bottleneck} of real-world data-intensive applications.
}

\textbf{To this end}, we \gagan{demonstrate the capability of near-HBM FPGA-based accelerator\gonur{s} for two key real-world data-intensive applications: (1) pre-alignment filtering in genome analysis, (2) representative kernels from \damlaa{a widely-used weather prediction application, COSMO}. 
\gmicro{Pre-alignment filtering is one of the fundamental steps in most genome \gcamera{analysis} \gonur{tasks,} where up to 98\% of \gonurnew{input genomic data is filtered out}. Thus, accelerating this step would benefit the overall end-to-end execution time of genome analysis~\cite{alser2019sneakysnake,senolcalimicro2020, alser2020accelerating, turakhia2018darwin,nag2019gencache,xin2013accelerating,georganas2015meraligner}. The weather kernels we evaluate are an essential part of climate and weather modeling \gonur{and prediction}~\cite{thaler2019porting}, which is critical for a  sustainable life ecosystem~\garxiv{\cite{schar2020kilometer}}.}

Our accelerator design\gonur{s} make use of a heterogeneous memory hierarchy (consisting \damlaa{of} URAM, BRAM, and HBM). We evaluate the performance and energy \damlaa{efficiency} of our accelerator\gonur{s,} perform a scalability analysis, and compare \gonur{them} to a traditional DDR4-based FPGA board and a state-of-the-art multi-core IBM POWER9 system.} 
\gmicro{Based on our analysis, we show that our full-blown HBM-based designs of \sneaky, \vadvc, and \hdiff provide (1)~$27.4\times$, $5.3\times$, and $12.7\times$ higher speedup, and (2)~$133\times$, $12\times$, and $35\times$ higher energy efficiency\src{, respectively,} compared to a 16-core IBM POWER9 system.}




\section{Near-memory Computation on \damlaa{FPGAs}}

\begin{figure}[h]
 \centering
 \includegraphics[width=\linewidth,trim={0.5cm 1.2cm 0.5cm 1cm},clip]{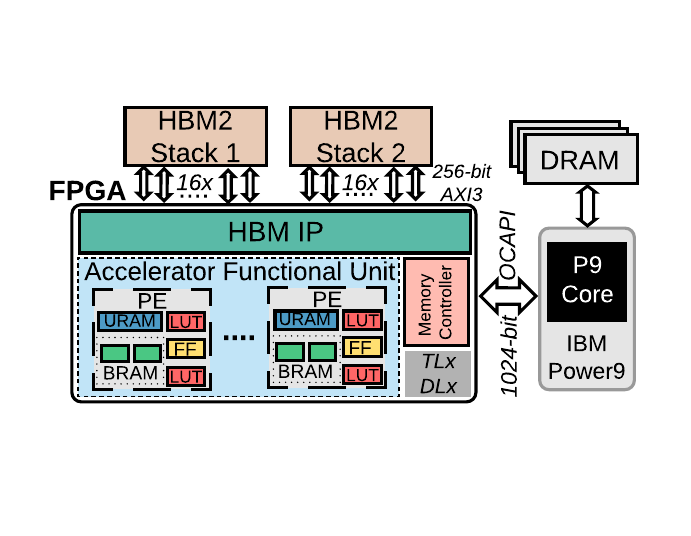}
 \vspace{-20pt}
 \caption{Heterogeneous platform with an IBM POWER9 system connected to an HBM-based FPGA board via~OCAPI. We also show components of an FPGA: flip-flop (FF), lookup table (LUT), UltraRAM (URAM), and Block RAM (BRAM).
 \label{fig:system}}
\end{figure}

\sr{We provide in} \juan{Figure~\ref{fig:system} a high-level schematic of our integrated system with \damlaa{an FPGA-based near-memory} accelerator. 
The FPGA is connected to two HBM stacks, each \sr{of which has} 16 \emph{pseudo memory channels}~\cite{axi_hbm}. A channel is exposed to the FPGA as a 256-bit wide interface, and the FPGA has 32 such channels {in total}. The HBM IP provides 8 memory controllers (per stack) to handle the data transfer \sr{to/from} the HBM memory channels. 
This configuration enables high-bandwidth and low-latency memory \damla{accesses} for near-memory computing. 
The FPGA is also connected to a host CPU, an IBM POWER9 processor, using \sr{OCAPI (OpenCAPI)}~\cite{openCAPI}.}

\juan{The FPGA device implements an \emph{accelerator functional unit} (AFU) that interacts with the host system through the  TLx (Transaction Layer) and the DLx (Data Link Layer), which are the CAPI endpoints on the FPGA. An AFU comprises multiple \emph{processing elements} (PEs) that accelerate \damla{a} \damlaa{portion} of an application.}


\section{Modern Data-Intensive Applications}

\subsection{\damla{Case Study 1: Pre-Alignment Filtering in Genome Analysis}}
One of the most fundamental computational steps in most \gcamera{genome analysis} \gonur{tasks} is sequence alignment~\damlaa{\cite{alser2020accelerating,alser2020technology}}.
This step is formulated as \alser{an} \emph{approximate string matching} (ASM) problem~\garxivCite{\cite{senolcalimicro2020,navarro2001guided}} and it calculates: (1)~\emph{edit distance} (the minimum number of edits needed to convert one sequence into the other) between two given sequences~\garxiv{\cite{levenshtein1966binary,navarro2001guided}}, (2)~type of each edit (i.e., insertion, deletion, or substitution), (3)~location of each edit in one of the two given sequences, and (4)~\emph{alignment score} that is the sum of the scores (\gonur{calculated using} a user-defined scoring function) of all edits and matches between the two sequences.

Sequence alignment is a computationally-expensive step as it usually uses dynamic programming~(DP)-based algorithms~\garxivCite{\cite{smith1981identification,needleman1970general,vsovsic2017edlib, daily2016parasail, li2018minimap2, senolcalimicro2020}}, which \gagan{have} quadratic time and space complexity (i.e., O($m^2$) for a sequence length of $m$).
In genome analysis, an overwhelming majority ($>$98\%) of the sequence pairs examined during sequence alignment are highly dissimilar and their alignment results are \emph{simply} discarded as such dissimilar sequence pairs are usually not useful for genomic studies~\cite{xin2013accelerating,xin2015shifted,alser2017gatekeeper}.
To avoid examining dissimilar sequences using computationally-expensive sequence alignment algorithms, genome analysis pipelines typically use filtering heuristics that are called \emph{pre-alignment filters}~\garxiv{\cite{alser2020accelerating, alser2017gatekeeper,alser2017magnet1,alser2019shouji,alser2019sneakysnake,bingol2021gatekeeper}}.
The key idea of pre-alignment filtering is to quickly estimate the number of edits between two given sequences and use this estimation to decide whether or not the computationally-expensive DP\aov{-based alignment} calculation is needed --- if not, a significant amount of time is saved by avoiding DP\aov{-based alignment}. 
If two genomic sequences differ by more than an edit distance threshold, $E$, then the two sequences are identified as dissimilar sequences and hence DP calculation is not needed. 

\begin{figure}[h]
 \centering
 \includegraphics[width=\linewidth]{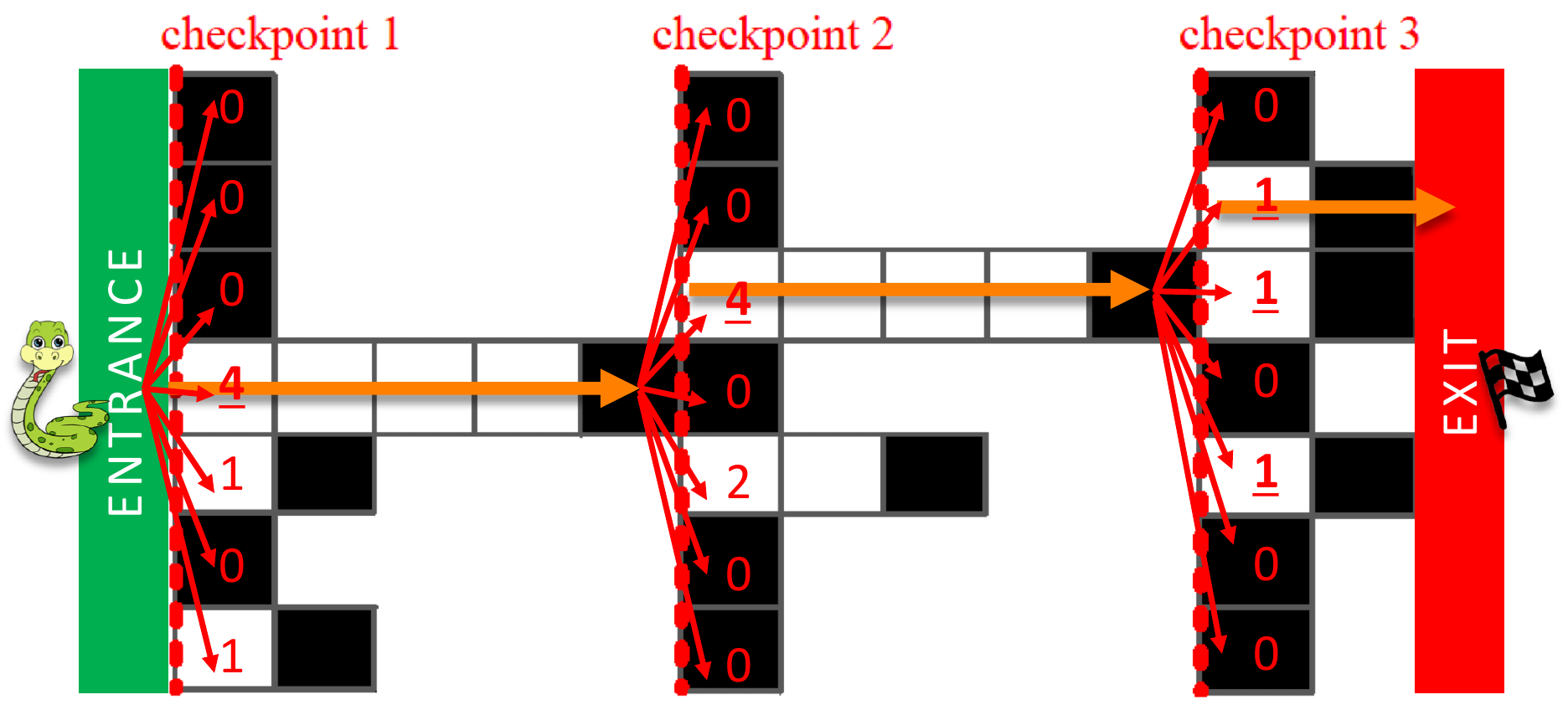}
 \vspace{-15pt}
  \caption{An example of the \gonur{\sneaky} chip maze for a reference sequence $R$ = ‘GGTGCAGAGCTC’, a query sequence 
  $Q$= ‘GGTGAGAGTTGT’, and an edit distance threshold ($E$) of 3. 
  Our \sneaky algorithm quickly finds an optimal signal net (highlighted in orange) with 3 obstacles, each of which is located at the end of each arrow (subpath), and hence \sneaky decides that sequence alignment for $R$ and $Q$ is needed, as the number of obstacles $\leq$ $E$.}
 \label{fig:SneakySnake}
 \end{figure}

\emph{\sneaky}~\cite{alser2019sneakysnake} is a recent highly-parallel and highly-accurate pre-alignment filter that works on {\emph{modern} high-performance computing architectures such as CPUs, GPUs, and FPGAs}. 
The key idea of \sneaky is to 
reduce the ASM problem to the \emph{single net routing} (SNR) problem~\cite{lee1976use} \damlaa{in VLSI}. 
The \damla{goal of the} SNR problem is to find the shortest routing path that interconnects two terminals on the boundaries of VLSI chip layout {while passing} through the minimum number of obstacles.
Solving the SNR problem is faster than solving the ASM problem, as calculating the routing path after facing an obstacle is independent of the calculated path before this obstacle (checkpoints in Figure~\ref{fig:SneakySnake}).
This provides two key benefits\damla{: 1)~It} obviates the need for using computationally-costly DP algorithms to keep track of the subpath that provides {the} optimal {solution} (i.e., {the one} with the least possible routing cost).
2)~The independence {of} the subpaths allows for solving many SNR subproblems in parallel by judiciously leveraging the parallelism-friendly architecture of modern FPGAs and GPUs to greatly speed up the \sneaky algorithm.
The number of obstacles faced throughout the found routing path represents a \emph{lower bound} on the edit distance between two sequences\damlaa{, and hence this number is used to make accurate filtering decisions by \sneaky.}
\newline
The \sneaky algorithm \damla{includes} three main steps~\cite{alser2019sneakysnake}:
1)~\textbf{Building the chip maze}. The chip maze, $Z$, is a matrix where each of its entries represents the pairwise comparison result of a character of one sequence with another character of the other sequence\gonur{, as we show in Figure~\ref{fig:SneakySnake}}.
Given two genomic sequences \gagan{of length $m$}, a reference sequence $R[1\dots m]$ and a query sequence $Q[1\dots m]$, and an edit distance threshold $E$, \sneaky calculates the entry $Z[i, j]$ of the chip maze as follows:\newline
\begin{equation}
\label{equ:chipmaze}
\small{\left\{\begin{matrix*}[l]
0,\:\: if~i=E+1, ~Q[j]=R[j], \\ 
0,\:\: if~1\leq i\leq E, ~Q[j-i]=R[j], \\ 
0,\:\: if~i> E+1, ~Q[j+i-E-1]=R[j], \\ 
1,\:\: otherwise 
\end{matrix*}\right.} \newline
\end{equation}
where an entry of value `1' represents an obstacle, an entry of value `0' represents an available path, $1\leq i\leq (2E+1)$, and $1\leq j\leq m$.

2)~\textbf{Finding the longest available path in each row of the chip maze}. \sneaky finds the longest available path, which represents the longest common subsequence between two given sequences.
It counts the consecutive entries of value 0 in each row starting from the previous checkpoint until it faces an obstacle, \gonur{and} then it examines the next rows in the same way.
Once it examines all rows, \gmicro{\sneaky compares the lengths of the found segments of consecutive zeros and considers the longest segment as the chosen path \gonur{(arrows highlighted in orange in Figure~\ref{fig:SneakySnake})}, and places a checkpoint right after the obstacle that follows the \damla{chosen} path. \sneaky then starts the count from this checkpoint for each row of the chip maze. Thus, the CPU-based implementation consists of irregular memory access patterns. }

3)~\textbf{Finding the estimated number of edits}. \gonur{\sneaky estimates the number of edits to be equal to the number of obstacles found along the shortest routing path (3 obstacles for our example in Figure~\ref{fig:SneakySnake}, each of which is located at the end of each chosen subpath).
Thus,} \sneaky repeats the second step until either \damlaa{there are} no more available entries to be examined, or the total number of obstacles passed through exceeds the allowed edit distance threshold (i.e., $E$).


\subsection{\damla{Case Study 2: Weather \gonur{Modeling and} Prediction}}

\juan{Accurate \gagan{and fast} weather prediction using detailed weather models is essential to make weather-dependent decisions in a timely manner. 
The Consortium for Small-Scale Modeling (COSMO)~\cite{doms1999nonhydrostatic} built one such weather model to meet the high-resolution forecasting requirements of weather services. 
The COSMO model is a non-hydrostatic atmospheric prediction model \sr{that is widely used} for meteorological purposes and research applications~\garxiv{\cite{schar2020kilometer,gysi2015modesto}}}.

\juan{The central part of the COSMO model (called \emph{dynamical core} or \emph{dycore}) solves the Euler equations on a curvilinear grid~\garxiv{\cite{wicker2002euler}} and applies \damla{1)~}implicit discretization (i.e., parameters are dependent on each other at the same time instance~{\garxivCite{\cite{bonaventura2000semi}}}) in the vertical dimension and \damla{2)~}explicit discretization (i.e., a solution \sr{depends} on the previous system state~{\garxiv{\cite{bonaventura2000semi}}}) in the \gagan{two} horizontal dimensions. The use of different discretizations leads to three computational patterns~{\cite{cosmo_knl}}: 1)~horizontal stencils, 2)~tridiagonal solvers in the vertical dimension, and 3)~point-wise computation. 
These computational kernels are compound stencil kernels that operate on a three-dimensional grid~\garxiv{\cite{gysi2015modesto, singh2019low}}. A stencil operation updates values in a structured multidimensional grid (\emph{row}, \emph{column}, \emph{depth}) based on the values of a fixed local neighborhood of grid points. 
\emph{Vertical advection} (\texttt{vadvc}) and \emph{horizontal diffusion} (\texttt{hdiff}) are \gcamera{two} such compound stencil kernels found in the \emph{dycore} of the COSMO weather prediction model. They are similar to the kernels used
in other weather and climate models~\garxiv{\cite{kehler2016high,neale2010description,doi:10.1175/WAF-D-17-0097.1}}.

These kernels are representative of the data access patterns and algorithmic complexity of the entire COSMO model, and are similar to the kernels used in other weather and climate models~\cite{singh2020nero}. 
As shown in Figure~\ref{fig:roofline}, their performance is \gagan{bounded} by access to \gonur{main} memory due to their irregular memory access patterns and low arithmetic intensity that \gonur{altogether} \damla{result} in an order-of-magnitude lower performance than the peak CPU performance.} 

\begin{figure}[h]
\centering
    \resizebox{0.9\textwidth}{!}{

    \begin{tikzpicture}[rotate=90,transform shape]

    \gridThreeD{0}{8.25}{black};
    \node at (8,10.5)[ rotate=270] {\huge \textit{Laplacian}};
    \node at (7.35,10.5)[ rotate=270] {\huge \textit{Stencil}};
      \dotesUpper{0}{8.25};
    \gridThreeDSecond{1}{4}{black};
    \node  at (8,5.5)[ rotate=270] {\huge \textit{Flux}};
    \node at (7.35,5.5)[ rotate=270] {\huge \textit{Stencil}};
     \lowerDots{3}{4};
     
    \gridThreeDThird{2}{0}{black};
     \node  at (7.5,0.25) [ rotate=270] {\huge \textit{Output}};
     \lowestDot{1.6}{-0.45}

      \draw [thick,->,blue] (2.75,4.25) -- (2.7,8.4);
       \draw [thick,->,blue] (2.75,4.25) -- (2.1,8.9);
       \draw [thick,->,blue] (2.75,4.25) -- (4.1,8.9);
       \draw [thick,->,blue] (2.75,4.25) -- (3.5,9.4);

       \draw [thick,->,orange] (2.1,4.7) -- (3.5,9.4);
       \draw [thick,->,orange] (2.1,4.7) -- (1.5,9.4);
       \draw [thick,->,orange] (2.1,4.7) -- (2.1,8.9);
       \draw [thick,->,orange] (2.1,4.7) -- (2.9,9.9);
     
       \draw [thick,->,red] (4.1,4.7) -- (3.5,9.4);
       \draw [thick,->,red] (4.1,4.7) -- (5.5,9.4);
       \draw [thick,->,red] (4.1,4.7) -- (4.1,8.9);
       \draw [thick,->,red] (4.1,4.7) -- (4.9,9.9);

           \draw [thick,->,brown] (3.5,5.25) -- (3.5,9.4);
           \draw [thick,->,brown] (3.5,5.25) -- (2.9,9.9);
           \draw [thick,->,brown] (3.5,5.25) -- (4.9,9.9);
           \draw [thick,->,brown] (3.5,5.25) -- (4.3,10.4);
         \draw [thick,->,violet] (2.7,0.25) -- (3.5,5.15);
         \draw [thick,->,violet] (2.7,0.25) -- (4.1,4.65);
         \draw [thick,->,violet] (2.7,0.25) -- (2.1,4.65);
         \draw [thick,->,violet] (2.7,0.25) -- (2.75,4.15);

    

\end{tikzpicture}
}

\caption{Horizontal diffusion kernel composition \gonur{using Laplacian and flux stencils} in a two dimensional plane~\cite{singh2019narmada}.
\label{fig:hdiffMemory}}
\vspace{-0.2cm}
\end{figure}
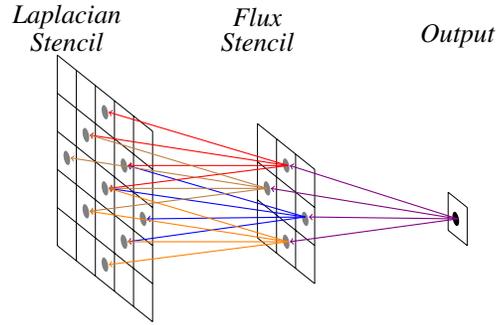

The horizontal diffusion kernel iterates over a 3D grid performing \textit{Laplacian} and \textit{flux} \gonur{stencils} to calculate different grid points as shown in Figure~\ref{fig:hdiffMemory}. \gmicro{A single \textit{Laplacian} stencil accesses
the input grid at five memory offsets, the result of which is used to calculate the \textit{flux} \gonur{stencil}. \hdiff has purely horizontal access patterns and does not have dependencies in the vertical dimension. \damlaaa{Thus,} it can be fully parallelized in the vertical dimension.}

Vertical advection has a higher degree of complexity since it uses the Thomas algorithm~\cite{thomas} to solve a tridiagonal matrix of \gonur{weather data (called \emph{fields}, such as, air pressure, wind velocity, and temperature)} along the vertical axis. \vadvc consists of a forward sweep that is followed by a backward sweep along the vertical dimension. \vadvc requires access to \gonur{the weather data,} which are stored as array structures while performing forward and sweep computations.
Unlike the conventional stencil kernels, vertical advection has dependencies in the vertical direction, which leads to limited available parallelism and irregular memory access patterns. \gmicro{ For example, when the input grid is stored by \emph{row}, accessing data elements in the \emph{depth} dimension typically results in \gonur{many} cache \gonur{misses}~\cite{xu2018performance}. }

\section{Accelerator Implementation}

\gagan{We design and implement an accelerator on \gonur{our} HBM-based FPGA-board (Figure~\ref{fig:system}) for each of the \juan{three kernels (\sneaky, \vadvc, and \hdiff) in our} two case studies. \gmicro{We use a High-Level Synthesis (HLS)~\garxiv{\cite{hls}} design flow to implement and map our accelerator design.} \gonur{We} 
\juan{describe the design and the execution flow for our HBM-based accelerators.}}


\begin{figure*}[h]
 \centering
 \includegraphics[width=\textwidth,trim={0.45cm 0.5cm 0.5cm 0.5cm},clip]{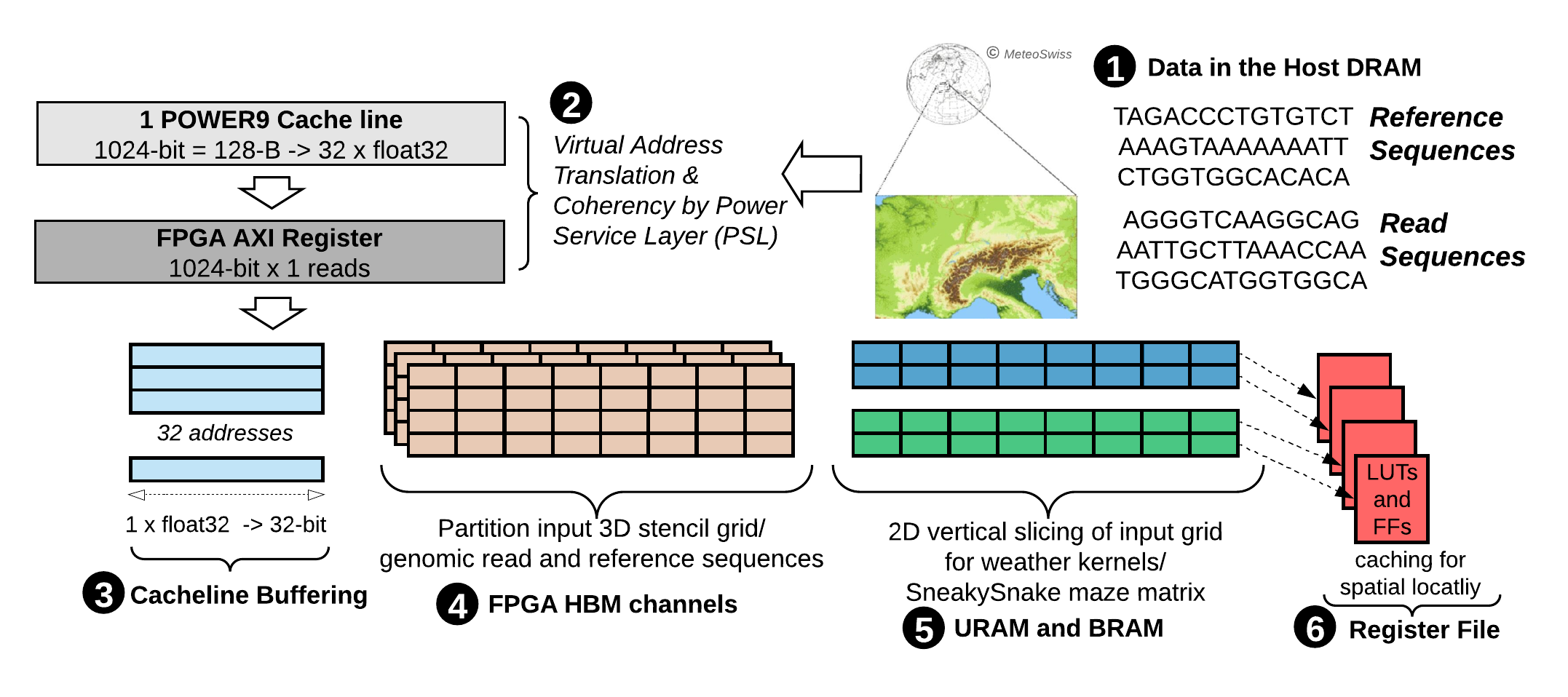}
 \vspace{-20pt}
 \caption{\gonur{Data transfer flow} from the host DRAM to the on-board FPGA memory via POWER9 \src{cache lines}. Heterogeneous memory partitioning of on-chip
memory blocks enable \gonur{low read/write} latenc\gonur{ies} across the FPGA memory hierarchy.
 \label{fig:dataflow_seq}}
\end{figure*}

Figure~\ref{fig:dataflow_seq} shows the end-to-end data transfer from the host DRAM to the processing element on an FPGA. \gmicro{We make use \did{of} streams (\texttt{hls::streams}\footnote{We \gonur{use} Vivado HLS \did{C++ template class \textit{hls::stream}} to implement \did{FIFO-based} streaming interfaces.}) to connect different dataflow tasks that allow consumer functions to operate before the producer functions have been completed. Streaming simplifies address management as the data samples are sent in sequential order between two modules. Before feeding data to a processing element, we use the on-chip heterogeneous memory hierarchy to unpack the stream data in \gonur{a way that matches} the data access pattern of an application.} Therefore, we implement our accelerator design following a dataflow approach \sr{in five steps}. 

First, the input data stored in the DRAM of the host system (\circled{1} in Figure~\ref{fig:dataflow_seq}) is transferred to the FPGA over \gmicro{a 1024-bit wide OCAPI interface (\circled{2}) by a \textit{data-fetch engine}. \gonurnew{A single cache line stream of \texttt{float32} datawidth would have 32 data elements.} The data-fetch engine reads 1024-bit wide POWER9 \src{cache line} data over the OCAPI interface and pushes the data into a 1024-bit buffer before converting it to 256-bit HBM pseudo-channel bitwidth. } \damlaa{For} weather prediction, the input data is the atmospheric data collected from weather simulations based on the atmospheric model resolution grid. \damlaa{For} genome analysis, the input data \gonur{is the} reference and read sequences for the pre-alignment filtering \damla{step of the genome analysis pipeline}. Second, \gmicro{following the initial buffering (\circled{3}),} \gonur{the} \textit{HBM-write engine} maps the data onto the HBM memory (\circled{4}). \gagan{We partition the data among HBM channels (\circled{5}) to exploit data-level parallelism and \damlaa{to scale} our design. Our evaluated workloads have limited locality, so to exploit locality, we cache certain parts of data into a register file \gonur{made of LUTs and FFs} (\circled{6}).} Third, we assign a dedicated \damla{HBM} memory channel to a specific \damla{processing element (PE)}; therefore, we enable as many HBM channels as the number of PEs. \gmicro{This allows us to use the high HBM bandwidth effectively because each PE fetches from \damlaaa{an} independent 256-bit channel, which provides low-latency \gonur{and high-bandwidth} data access to each PE.} An \textit{HBM-read engine} reads data from a dedicated \damla{HBM} memory channel and assigns data to a specialized PE. \gmicro{The HBM
channel provides 256-bit data, which is a quarter of \gonur{the} OCAPI
bitwidth (1024-bit). Therefore, to match the OCAPI bitwidth,
we introduce a stream converter logic that converts a
256-bit HBM stream to a 1024-bit wide stream, which is equal to the maximum OCAPI bitwidth.}

Fourth, each PE performs computation (\sneaky ~\damlaa{for} genome analysis, and \vadvc or \hdiff ~\damlaa{for} weather prediction) on the transferred data. \gmicro{In \sneaky, we equally divide the number of \gonur{read} and reference sequences among the PEs. \gonur{In} \vadvc and \hdiff, each PE operates on a block of the input grid.} \gmicro{For \sneaky\footnote{\gonurnew{We open-source our \texttt{HBM+OCAPI}-based \sneaky accelerator implementations (both single-channel-single PE and multi-channel-single PE): \MYhref[black]{https://github.com/CMU-SAFARI/SneakySnake/tree/master/SneakySnake-HLS-HBM}{https://github.com/CMU-SAFARI/SneakySnake/tree/master/SneakySnake-HLS-HBM}}}, each row in the chip maze is stored as a register array of \src{length} \gonur{equal} to the read length. The registers are accessed simultaneously throughout the execution. In every iteration, we count \src{consecutive} zeros in each row until \src{we find} an obstacle (i.e., until we come across a 1). Following this, we shift all the bits by the max\gonur{imum} number of zeros in the chip maze. This shifting allows us to overcome the irregular array accesses while finding the longest possible path in the chip maze.} Fifth, \gonur{once the calculated results are available,} the \textit{HBM-write engine} writes calculated results to its assigned HBM memory channel, \gonur{after} which \gonur{the} \textit{write-back engine} transfers the data back to the host system for further processing. 

\gagan{We create a specialized memory hierarchy from the \did{pool of } heterogeneous FPGA memories \gagan{(i.e., \did{on-chip} BRAM and URAM, and \did{in-package} HBM)}. By using a greedy algorithm, we determine the best-suited hierarchy for each kernel. Heterogeneous partitioning of on-chip memory blocks reduces read and write latencies across the FPGA memory hierarchy. }To optimize a PE, we apply \sr{three} optimization strategies. First, we exploit the inherent parallelism in a given algorithm \gonur{using} hardware pipelining. Second, we partition  \gonur{data arrays onto multiple physical on-chip memories (BRAM/URAM) instead of a single large memory to avoid} stalling of our pipelined design, since the on-chip BRAM/URAM have only two read/write ports. \did{On-chip memory reshaping is an effective technique for improving the bandwidth of BRAMs/URAMs.}
\gagan{Third, we partition the input data between PEs, therefore, all PEs exploit data-level parallelism.} \did{\gonur{We apply all} three optimizations \gonur{using} source-code annotations \gonur{via}~\textit{Vivado~HLS~\#pragma} directives~\garxiv{\cite{hls_pragma}}}.


\section{Evaluation}
\label{section:evaluation}
\gcamera{We evaluate our accelerator designs for \sneaky, \vadvc, and \hdiff in terms of performance, energy consumption, and FPGA resource utilization on two different FPGAs, and two different external data communication interfaces between the CPU and the FPGA board.}
\gagan{We implement our accelerator designs 
on \sr{both 1)} \gonur{an} Alpha-Data ADM-PCIE-9H7 card~\garxiv{\cite{ad9h7} }featuring the Xilinx Virtex Ultrascale+ XCVU37P-FSVH2892-2-e~\garxiv{\cite{vu37p}} with 8GiB HBM2~\cite{hbm} \sr{and 2)} \gonur{an} Alpha-Data ADM-PCIE-9V3 card~\garxiv{\cite{ad9v3}} featuring the Xilinx Virtex Ultrascale+ XCVU3P-FFVC1517-2-i with 8GiB DDR4~\garxiv{\cite{vu37p}}, connected to an IBM POWER9 host system. For the external data communication interface, we use both CAPI2~\garxiv{\cite{stuecheli2015capi}} and the state-of-the-art OCAPI (OpenCAPI)~\garxiv{\cite{openCAPI}} interface. 
We compare these implementations to execution on a POWER9 CPU with 16 cores \juan{(\damlaa{using all} 64 hardware threads).} 
\sr{We run \sneaky using the first \gonur{30,000} real genomic sequence pairs (text and query pairs) of \texttt{100bp\_2} dataset~\cite{alser2019sneakysnake}, which is widely used as in prior works~\garxiv{\cite{kim2018grim,georganas2015meraligner,alser2019shouji,alser2017gatekeeper,bingol2021gatekeeper}}.
The length of each sequence is 100 bp (base-pair) long.} For weather prediction, we run our experiments using a $256\times256\times64$-point domain similar to the grid domain used by the COSMO model. }

\subsection{Performance Analysis}
\label{subsection:perf}
\begin{figure*}[t]

 \begin{subfigure}{\textwidth}
  \includegraphics[width=\textwidth,trim={0cm 0cm 0cm 0cm},clip]{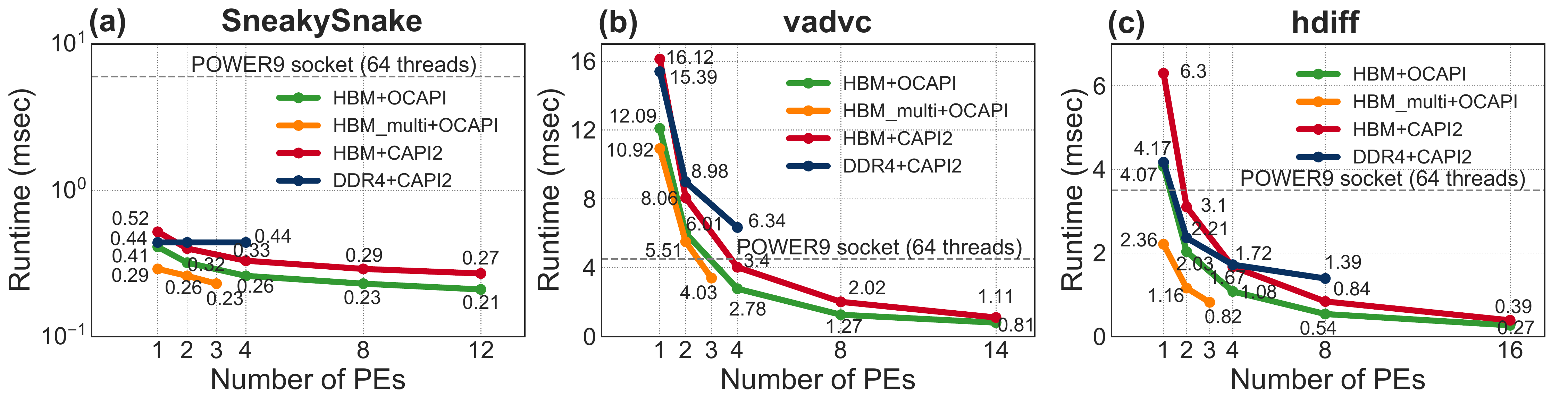} 
 \end{subfigure}
 \begin{subfigure}{\textwidth}
  \hspace{0.1cm}
 \includegraphics[width=1\linewidth]{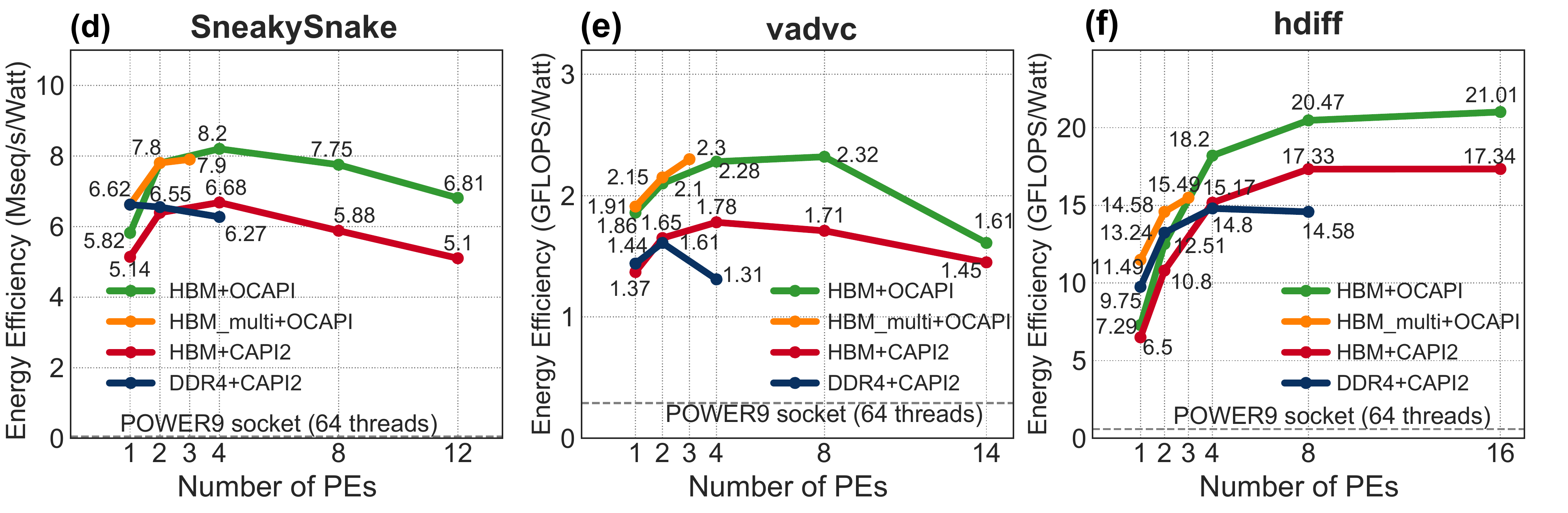}
 \end{subfigure}
\vspace{-8pt}
 \caption{Performance for (a) \sneaky, (b) \vadvc, and (c) \texttt{hdiff} as a function of accelerator PE count on the HBM- and DDR4-based FPGA boards. Energy efficiency for (d) \sneaky, (e) \texttt{vadvc}, and (f) \texttt{hdiff} and on HBM- and DDR4-based FPGA boards. 
 We also show the single socket (64 threads) performance and energy efficiency of an IBM POWER9 host system for \sneaky, \texttt{vadvc}, and \texttt{hdiff}. \gmicro{For HBM-based design, we implement our accelerator with \gonur{both the} CAPI2 interface and the state-of-the-art OpenCAPI (OCAPI) interface (with both single channel and multiple \sr{channels} per PE).} 
 }
 \label{fig:results}
 
\end{figure*}

\sr{We provide the execution time of \sneaky, \vadvc, and \hdiff on the POWER9 CPU with 64 threads and the FPGA accelerators (both DDR4-based and HBM-based) in Figure~\ref{fig:results} (a), (b), and (c), respectively.}
\juan{
For both FPGA designs, we scale the number of PEs from 1 to the maximum number that we can accommodate on the available FPGA resources. 
On the DDR4-based design, the maximum number of PEs is 4, 4, and 8 for \sneaky, \vadvc, and \hdiff , respectively. 
On the HBM-based design, we can fit up to 12, 14, and 16 PEs for \sneaky, \vadvc, and \hdiff, respectively. 
\damla{Based on our analysis, we make \textbf{four key observations.}}}


First, the full-blown HBM+OCAPI-based implementations (with the maximum number of PEs) of \sr{\sneaky, \vadvc}, and \hdiff { outperform the 64-thread \damlaa{IBM POWER9} CPU version by $27.4\times$, $5.3\times$, and $12.7\times$\sr{, respectively}. We achieve 28\%, 37\%, and 44\% higher performance for \sneaky, \vadvc, and \hdiff, respectively, with OCAPI-based HBM design than CAPI2-based HBM design \gonur{due to} the following two reasons: 1) OCAPI provides double the bitwidth (1024-bit) of the CAPI2 interface (512-bit), which provides a higher bandwidth to the host CPU, \did{i.e., 22.1/22.0 GB/s R/W versus 13.9/14.0 GB/s}; and 2) with OCAPI, memory coherency logic \gonur{is} moved onto the IBM POWER CPU, which provides more FPGA area and allows us to run our accelerator logic at a higher clock \gonur{frequency} \gcamera{(250MHz for OCAPI \gonur{versus} 200MHz for CAPI2).}}

\juan{Second, for a single PE, the DDR4-CAPI2-based \damlaa{FPGA accelerator} design is faster than the HBM-CAPI2-based design for all three kernels. 
\damlaa{This is because} the HBM-based design uses one HBM channel per PE\gonur{, as} the bus width of the DDR4 channel (512 bits) is larger than that of an HBM channel (256 bits). 
\gagan{ Therefore, the HBM channel has a lower transfer rate of 0.8-2.1 GT/s (Gigatransfers per second) than for a DDR4 channel (2.1-4.3 GT/s) \gmicro{with a theoretical bandwidth of 12.8 GB/s 
and 25.6 GB/s 
per channel, respectively.}}} One way to match the DDR4 bus width 
\juang{is} to have a single PE fetch data from multiple HBM \damlaaa{channels} in parallel. \gmicroF{As shown in Figure~\ref{fig:results}, in our \damlaaa{multi-channel} setting (\gonur{\texttt{HBM\_multi+OCAPI}}), we use 4 HBM pseudo channels per PE to meet the bitwidth of the OCAPI interface. We observe that by fetching more data from multiple \damlaaa{channels}, compared to our single-\damlaaa{channel-}single PE design, we achieve $1.4\times, 1.2\times,$ and $1.8\times$ performance \damlaaa{improvement}  for \sneaky, \vadvc, and \hdiff, respectively.}

\juan{Third, as we increase the number of PEs, we divide the workload evenly across PEs. As a result, we observe linear scaling in \sr{the performance of} HBM-based designs, where each PE reads and writes through 
\damla{a dedicated} HBM channel. \gmicroF{\damlaaa{For multi-channel designs,} we \damlaaa{are} able to accommodate only 3 PEs \gcamera{for the three evaluated kernels} (i.e., 12 HBM channels) because adding more HBM channels leads to \gcamera{timing constraint violations}. We observe that \damlaaa{the} 
\juang{best-}performing \damlaaa{multi-channel-single PE} design (i.e., using 3 PEs with 12 HBM channels \juang{for \damlaaa{all} three workloads}) \src{has} $1.1\times$, $4.7\times$, \src{and} $3.1\times$ \gonur{lower performance than the} 
\juang{best-}performing \damlaaa{single-channel-single PE design} \juang{(i.e., 12 PEs for \sneaky, 14 PEs for \vadvc, \gonur{and} 16 PEs for \hdiff\src{, respectively})}. 
\juang{This observation shows that there is a tradeoff between (1)~enabling more HBM pseudo channels to provide each PE with more bandwidth, and (2)~implementing more PEs in the available area.}
\gonur{For} \sneaky, \gonur{in our dataflow design,} \gonur{data transfer} time dominates the computation \gonur{time}; therefore, adding more PEs \damlaaa{does} not lead to a linear reduction in performance. \damlaaa{\gonur{For} \vadvc and \hdiff}, both \gonur{data transfer} and computation take a comparable amount of time. Therefore, in such workloads, we \gonur{are able to achieve a linear execution time reduction with the number of PEs.} } }

\juan{Fourth, \sr{the performance of the DDR\damlaa{4}-based designs scales non-linearly for \vadvc and \hdiff with the number of PEs, as all PEs access memory through the same channel. Multiple PEs compete for a single memory channel, which causes frequent memory stalls due to} \gonur{contention} in the memory channel. 
For \sneaky, which is the most memory-bound of the three kernels, we observe a constant execution time with the increase in PEs. With a single PE, memory access time hides all computation time. Increasing the number of PEs reduces the time devoted to computation, but not \damla{the} memory access time because there is a single channel. Therefore, memory bandwidth saturates with a single PE, and the total execution time does \damla{\emph{not}} reduce with the number of PEs.}

\sr{We conclude that \damlaa{FPGA-based \gmicro{near-memory} \gcamera{computing} provides significant speedup (\gonur{between 5.3$\times$-27.4$\times$}) to key data-intensive applications over a state-of-the-art CPU (POWER9).}} 

\subsection{Energy \gcamera{Efficiency} Analysis}
\src{We provide energy efficiency results for \sneaky, \vadvc, and \hdiff on the two FPGA designs and the POWER9 CPU in Figure~\ref{fig:results} (d), (e), and (f), respectively}. 
\juan{We express energy efficiency in Mseq/s/Watt (i.e., millions of read sequences per second per Watt) for \sneaky, and in terms of GFLOPS/Watt for \hdiff and \vadvc. 
For power measurement on the POWER9 system \gagan{with an FPGA board}, we use the \gcamera{AMESTER tool\footnote{\MYhref[black]{https://github.com/open-power/amester}{https://github.com/open-power/amester}}} to monitor built-in power sensors. We measure the active power consumption, i.e., the difference between the total power of a complete socket (including processor, memory, fans, and I/O) when running an application and when idle.} 
\damla{Based on our analysis, we make \textbf{five key observations}.}

\juan{First, our full-blown HBM+OCAPI-based accelerator designs (with 12 PEs for \sneaky, 14 PEs for \vadvc, and 16 PEs for \hdiff ) \gagann{improve} energy efficiency \gagann{by} \gmicro{$133\times$, $12\times$, and $35\times$ compared to \gonur{the} POWER9 system for \sneaky, \vadvc, and \hdiff, respectively.}} 

\juan{Second, the DDR4-CAPI2-based  designs are \gagann{slightly} more energy efficient \gagann{($1.1\times$ to $1.5\times$)} than the HBM-CAPI2-based designs when the number of PEs is small. This observation is \sr{in line} with our discussion about performance with small PE counts in the previous section. 
However, as we increase the number of PEs, the HBM-based designs provide higher energy efficiency since they make use of multiple HBM channels.}
\gmicroF{Third, compared to our \damlaaa{single-channel-single PE design}, our \damlaaa{multi-channel-single PE design} provides only $1.1\times$, $1\times$, \src{and} $1.5\times$ \damlaaa{higher energy efficiency} for \sneaky, \vadvc, and \hdiff, respectively. \damlaaa{This is because using more channels leads to higher power consumption ($\sim$1 Watt per channel) even though we get higher bandwidth per PE by using multiple channels.}}

\juan{Fourth, the energy efficiency of the HBM-based design for \hdiff increases with the number of PEs until a saturation point (8 PEs). 
\gonur{However, the energy efficiency of \sneaky and \vadvc (\texttt{HBM+CAPI2}) designs 
decreases after using more than 4 PEs, and that of the  \vadvc (\texttt{HBM+OCAPI}) design decreases after using more than 8 PEs.} 
 \gonur{This is because every additional HBM channel increases power consumption by $\sim1$ Watt (for the HBM AXI3 interface operating at 250MHz with a logic toggle rate of $\sim12.5$\%).} In case of \vadvc,  there is a large amount of control flow \sr{that} leads to large \gagan{and inefficient} resource consumption when increasing the PE count. This causes a high increase in power consumption with~14~PEs}.

\damlaa{Fifth, as we increase the number of PEs, performance of \sneaky increases, whereas energy efficiency may not follow the same trend. This is because
\sneaky spends significant amount of execution time in fetching data from memory, as mentioned in the discussion on performance analysis. Thus, this may lead to a reduction in energy efficiency, depending upon the number of HBM channels used.}

We conclude that \damlaa{increasing the number of \gonur{PEs and} enabled HBM channels does \emph{not} always increase energy efficiency.} 
However, data-parallel kernels like \texttt{hdiff} can achieve much higher performance in an energy-efficient manner with more PEs and HBM~channels. 

\subsection{FPGA Resource Utilization}

We list the resource utilization of \sneaky, \vadvc, and \hdiff on the \juan{FPGA board with HBM memory} in Table~\ref{tab:utilization}. 
We \damlaa{make} three observations. 
\juan{First, BRAM \gagan{utilization} is significantly higher than \gonur{utilization} of other resources. The reason is that we use \texttt{hls::streams} to implement input and output to different functions. Streams are FIFOs, which are implemented with BRAMs.} 
Second, \sneaky performs all computations  using flip-flops (FF) and lookup table registers (LUT).
\juan{It} does not require digital signal processing units (DSP) \juan{since it does not execute \damlaa{floating-point} operations}. 
\juan{Third, resource consumption of \vadvc is much higher than that of \hdiff because it has higher computation complexity and requires a larger number of \gagan{input parameters} for the compound stencil computation. 
We observe that there are enough resources to accommodate more than 16 PEs for \hdiff, but in this work, we use only a single HBM stack \gonur{due to} timing constraint violations. Therefore, in this work, the maximum number of PEs is 16 to match 16 memory channels offered by~a~single~HBM stack.}



\begin{table}[h]
\begin{center}
  \caption{FPGA resource utilization in the full-blown HBM+OCAPI-based designs for  \sneaky~(12 PEs), \vadvc (14 PEs), and \hdiff (16 PEs).}
\label{tab:utilization}
\resizebox{\linewidth}{!}{%
\begin{tabular}{llllllc}
\hline
\textbf{Algorithm} & \textbf{BRAM} & \textbf{DSP} & \textbf{FF} & \textbf{LUT} & \textbf{URAM} \\ \hline
\sneaky& 58\%            & 0\%            & 18\%           & 70\%           & 1\%  \\
\texttt{vadvc}              & 90\%            & 39\%           & 37\%          & 55\%           & 53\%                         \\ 
\texttt{hdiff}              & 96\%            & 4\%            & 10\%           & 15\%           & 8\%                  
                         \\ \hline
\end{tabular}
}
  \end{center}
\vspace{-0.2cm}
\end{table}
 
\section{Discussion}
\juan{This paper presents our \gonur{recent} efforts to leverage \gonur{near-memory computing capable FPGA-based accelerators} \gagan{to accelerate} \gmicro{three major kernels taken from} two data-intensive applications: (1) pre-alignment filtering in genome analysis, \gonur{and} (2) horizontal diffusion and vertical advection stencils from weather \gagann{prediction}. 
We identify key challenges for \gonur{such} acceleration and provide solutions to them. 
We summarize the most important insights and takeaways as follows.}

\juan{First, our evaluation shows that \gonur{High-Bandwidth Memory}-based \gonur{near-memory} \gagann{FPGA} accelerator designs can improve performance by \gonur{ 5.3$\times$-27.4$\times$ and energy efficiency by 12$\times$-133$\times$ over} a high-end \gagann{16-core} \gonur{IBM} POWER9 CPU.}

\juan{Second, our HBM-based \gagann{FPGA accelerator} designs employ a dedicated HBM channel per PE. This avoids memory access congestion, which is typical in DDR4-based FPGA designs and ensures memory bandwidth scaling with the number of PEs. As a result, \gagann{in most of the data-parallel applications,} performance scales linearly with the number of PEs.}

\juan{Third, \sr{the} maximum performance of our HBM-based design is \gonur{reached using} the maximum PE count that we can fit in the reconfigurable fabric, with each PE having a dedicated HBM channel.} \gagan{ However, adding more PEs could lead to \gcamera{timing constraint violation}\gonur{s} for HBM-based designs. HBM-based FPGAs consist of multiple super-logic regions (SLRs)~\garxiv{\cite{hbm_slr}}, where an SLR represents a single FPGA die. All HBM channels are connected only to SLR0, while other SLRs have indirect connections to the HBM channels. Therefore, for a large design, if a PE is implemented in a non-SLR0 region, it might make timing closure~difficult. }

Fourth, the energy efficiency of our HBM-based designs tends to saturate (or even \gonur{reduces}) as we increase the number of PEs \gonur{beyond some point}. The highest energy efficiency is achieved with a PE count that is smaller than the highest-performing PE count. The \gonur{major} reason \gonur{for a decrease in the energy efficiency is the increase in power consumption with} every additional HBM channel.


\gagann{We hope that \gonur{the} near-memory acceleration efforts for genome analysis and weather prediction \gonur{we described} and the challenges we identif\gonur{ied} and solve\gonur{d} \src{provide} a foundation for accelerating \gagann{modern} and future data-intensive applications \gagann{using powerful near-memory reconfigurable accelerators.}}

\section{ACKNOWLEDGMENTS}
\damlaa{\gcamera{We thank the \src{anonymous} reviewer\gonur{s} of IEEE Micro for their feedback. We thank the SAFARI Research Group members for valuable feedback and the stimulating intellectual environment they provide.} This work is supported by funding from \gcamera{\gonur{ASML,} Google, Huawei, Intel, Microsoft}\src{,} VMware, and the Semiconductor Research Corporation to Onur Mutlu.} Special thanks to IBM Research Europe, Zurich for providing access to IBM systems. This work was partially funded by Eindhoven University of Technology and ETH~Zürich.
\bibliographystyle{unsrtAlser}
\bibliography{ms}
\vspace{100pt}
\begin{IEEEbiography}{Gagandeep Singh}{\,}is with ETH Zürich, Zürich, Switzerland. Contact him at gagan.gagandeepsingh@safari.ethz.ch
\end{IEEEbiography}

\begin{IEEEbiography}{Mohammed Alser}{\,}is with ETH Zürich, Zürich, Switzerland. Contact him at alserm@ethz.ch
\end{IEEEbiography}

\begin{IEEEbiography}{Damla Senol Cali}{\,}is with Carnegie Mellon University,  Pittsburgh, PA, USA. Contact her at dsenol@andrew.cmu.edu
\end{IEEEbiography}

\begin{IEEEbiography}{Dionysios Diamantopoulos}{\,}is with IBM Research Europe, Zürich Lab, Rüschlikon, Switzerland. Contact him at did@zurich.ibm.com
\end{IEEEbiography}

\begin{IEEEbiography}{Juan G{\'{o}}mez-Luna}{\,}is with ETH Zürich, Zürich, Switzerland. Contact him at juan.gomez@safari.ethz.ch
\end{IEEEbiography}

\begin{IEEEbiography}{Henk Corporaal}{\,}is with Eindhoven University of Technology, Eindhoven, The Netherlands. Contact him at h.corporaal@tue.nl
\end{IEEEbiography}
\begin{IEEEbiography}{Onur Mutlu}{\,}is with {ETH Zürich, Zürich, Switzerland and Carnegie Mellon University, Pittsburgh, PA, USA.} Contact him at \gonur{omutlu@ethz.ch}
\end{IEEEbiography}



\end{document}